\documentclass[sigconf,screen,anonymous=false]{acmart}
\usepackage{amsmath}
\usepackage{bbm}
\usepackage{booktabs}
\usepackage{makecell}
\usepackage{balance}
\usepackage{multirow}

\newcommand{\norm}[1]{\left\lVert#1\right\rVert}

\usepackage{caption}

\usepackage{bbm}
\usepackage{figsize}
\usepackage{todonotes}

\newif\ifworkinprogress
\workinprogresstrue

\ifworkinprogress
    \newcommand{\HA}[1]{\textcolor{purple}{[Himan] #1}}
    \newcommand{\MM}[1]{\textcolor{red}{[Masoud] #1}}
  \newcommand{\RB}[1]{\textcolor{blue}{[Robin] #1}}
  \newcommand{\BM}[1]{\textcolor{orange}{[Bamshad] #1}}

\else
  \newcommand{\HA}[1]{}
  \newcommand{\MM}[1]{}
  \newcommand{\RB}[1]{}
  \newcommand{\BM}[1]{}
\fi

\AtBeginDocument{%
  \providecommand\BibTeX{{%
    \normalfont B\kern-0.5em{\scshape i\kern-0.25em b}\kern-0.8em\TeX}}}

\copyrightyear{2020}
\acmYear{2020}
\setcopyright{rightsretained}
\acmConference[RecSys '20]{Fourteenth ACM Conference on Recommender Systems}{September 22--26, 2020}{Virtual Event, Brazil}
\acmBooktitle{Fourteenth ACM Conference on Recommender Systems (RecSys '20), September 22--26, 2020, Virtual Event, Brazil}\acmDOI{10.1145/3383313.3418487}
\acmISBN{978-1-4503-7583-2/20/09}

\usepackage[absolute]{textpos}
\begin{document}
 \begin{textblock}{10}(3,0.4)
 \noindent\small  \begin{center}Accepted at the 14th ACM Conference on Recommender Systems (RecSys 2020) Late Breaking Results Track\end{center}
 \end{textblock}




\title{The Connection Between Popularity Bias, Calibration, and Fairness  in Recommendation}



\author{Himan Abdollahpouri}
\affiliation{%
  \institution{University of Colorado Boulder}
  \city{Boulder}
  \country{USA}}
\email{himan.abdollahpouri@colorado.edu}

\author{Masoud Mansoury}
\affiliation{%
  \institution{Eindhoven University of Technology}
  \city{Eindhoven}
  \country{Netherlands}
  }
  \email{m.mansoury@tue.nl}

\author{Robin Burke}
\affiliation{%
  \institution{University of Colorado Boulder}
  \city{Boulder}
  \country{USA}
  }
  \email{robin.burke@colorado.edu}

\author{Bamshad Mobasher}
\affiliation{%
  \institution{DePaul University}
  \city{Chicago}
  \country{USA}
  }
  \email{mobasher@cs.depaul.edu}

\renewcommand{\shortauthors}{Abdollahpouri et al.}

%
\begin{abstract}

Recently there has been a growing interest in fairness-aware recommender systems including fairness in providing consistent performance across different users or groups of users. A recommender system could be considered unfair if the recommendations do not fairly represent the tastes of a certain group of users while other groups receive recommendations that are consistent with their preferences. In this paper, we use a metric called miscalibration for measuring how a recommendation algorithm is responsive to users' true preferences and we consider how various algorithms may result in different degrees of miscalibration for different users. In particular, we conjecture that popularity bias which is a well-known phenomenon in recommendation is one important factor leading to miscalibration in recommendation. Our experimental results using two real-world datasets show that there is a connection between how different user groups are affected by algorithmic popularity bias and their level of interest in popular items. Moreover, we show that the more a group is affected by the algorithmic popularity bias, the more their recommendations are miscalibrated.

\end{abstract}

%
%

\maketitle              

\section{Introduction}

Recommendations are typically evaluated using measures such as precision, diversity, and novelty \cite{shani2011evaluating}. Under such measures, depending on the situation, a recommended list of items may be considered good if it is relevant to the user, is diverse, and also helps the user discover products that s/he would have not been able to discover in the absence of the recommender system.

One of the metrics used to measure recommendation quality is calibration, which measures whether the recommendations delivered to a user are consistent with the spectrum of items the user has previously rated. For example, if a user has rated 70\% action movies and 30\% romance, the user might expect to see a similar pattern in the recommendations \cite{steck2018calibrated}. If this ratio differs from the one in the user's profile, we say the recommendations are miscalibrated.

In addition to the mentioned metrics, recently there has been a growing interest in other aspects of the recommendations such as bias and fairness \cite{bozdag2013bias,dwork2012fairness}. There have been numerous attempts to define fairness \cite{narayanan2018translation,zafar2017fairness,singh2018fairness} and it is unlikely that there will be a universal definition that is appropriate across all applications. Recommendation fairness may have different meanings depending on the domain in which the recommender system is operating, the characteristics of different users or groups of users (e.g. protected vs unprotected), and the goals of the system designers. For instance, Mansoury et al. \cite{mansoury2020a} and Eksrand et al. \cite{ekstrand2018all} defined fairness as consistent performance across different groups of users. In their experiments, they observed certain groups such as females get less accurate recommendations than males. In addition, authors in \cite{DBLP:journals/corr/YaoH17} define several fairness metrics which focus on having a consistent performance in terms of estimation error across different user groups. In this paper, we use the same definition to measure (un)fairness. In other words, we consider an algorithm to be unfair if it does not provide consistent performance for different groups of users.

In this paper we focus on (mis)calibration as a quality of the recommendations. Miscalibration by itself may not be considered unfair as it could simply mean the recommendations are not personalized enough. However, if different users or groups of users experience different levels of miscalibration in their recommendations, this may indicate an unfair treatment of a group of users. Inspired by the work in \cite{ekstrand2018all,mansoury2020a} we are interested in investigating the potential factors that could lead to inconsistent performance of algorithms for different groups of users.

The class imbalance problem in machine learning typically is one of the main reasons for having unfair classification (inconsistent True Positive and Negative Rates across different classes). For instance, it is known that facial recognition systems have racial bias since different races are not equally represented in the training data \cite{garvie2016facial}. The equivalent of class imbalance in machine learning is popularity bias in recommender systems \cite{longtailnichesriche}: popular items are rated frequently, while the majority of other items do not get much attention. Recommendation algorithms are biased towards these popular items \cite{abdollahpouri2020multi}. We define algorithmic popularity bias as the tendency of an algorithm to amplify existing popularity differences across items. We measure this amplification through the metric of \textit{popularity lift}, which quantifies the difference between average item popularity in input (user profile) and output (recommendation list) for an algorithm.

It has been shown that popularity bias can lead to certain problems in recommendation such as shifting users' consumption towards more mainstream items over time and even causing homogenization of different groups of 
users \cite{mansoury2020b}. In this paper, we conjecture that popularity bias can be also an important factor leading to miscalibration of the recommendation lists for different users. We also show that users with different levels of interest in popular items get different levels of miscalibration. Our contributions are as follows:
\begin{itemize}

 \item \textbf{The disparate impact of popularity bias:} We show that different groups of users are affected differently by popularity bias based on how interested they are in popular items.
 \item \textbf{The connection between algorithmic popularity bias and miscalibration:} We show that users who are more affected by algorithmic popularity bias tend to also get less calibrated recommendations. 

\end{itemize}

\section{Related Work}
The problem of popularity bias and the challenges it creates for the recommender system has been well studied by other researchers \cite{anderson2006long,brynjolfsson2006niches,longtailrecsys}. Authors in the mentioned works have mainly explored the overall accuracy of the recommendations in the presence of long-tail distribution in rating data. In addition, some other researchers have proposed algorithms that can control this bias and give more chance to long-tail items to be recommended \cite{10.1109/TKDE.2011.15,DBLP:conf/recsys/KamishimaAAS14,flairs2019}. 

Moreover, the concept of fairness in recommendation has been also gaining a lot of attention recently \cite{kamishima2012fairness,DBLP:journals/corr/YaoH17}. For example, finding solutions that remove algorithmic discrimination against users belong to a certain demographic information \cite{zhu2018fairness} or making sure items from different categories (e.g. long tail items or items belong to different providers) \cite{liu2018personalizing} are getting a fair exposure in the recommendations. Our definition of fairness in this paper is aligned with the fairness objectives introduced by Yao and Huang in \cite{DBLP:journals/corr/YaoH17} where they define unfairness as having inconsistent estimation error across different users. We can generalize the \textit{estimation error} to simply be any kind of system performance such as the calibration of the recommendations as defined in \cite{steck2018calibrated}. In this paper, we use the same definition for fairness: a system is \textit{unfair} if it delivers different degree of miscalibration to different users. 

With regard to looking at the performance of the recommender system for different user groups, Ekstrand et al. \cite{ekstrand2018all} showed that some recommendation algorithms give significantly less accurate recommendations to groups from certain age or gender. In addition, Abdollahpouri et al. in \cite{abdollahpourimultistakeholder2020} discuss the importance of recommendation evaluation with respect to the distribution of utilities given to different stakeholders. For instance, the degree of calibration of the recommendation for each user group (i.e. a stakeholder) could be considered as its utility and therefore, a balanced distribution of these utility values is important for having a fair recommender system. 



\section{Popularity Bias and Miscalibration}
Popularity bias and miscalibration are both aspects of an algorithm's performance that are computed by comparing the attributes of the input data with the properties of the recommendations that are produced for users. In this section, we define these terms more precisely.

\subsection{Miscalibration}
One of the interpretations of fairness in recommendation is in terms of whether the recommender provides consistent performance across different users or groups of users. A recommender system could be considered unfair if the recommendations do not fairly represent the tastes of a certain group of users while other groups receive recommendations that are consistent with their preferences. In this paper we use a metric called miscalibration \cite{steck2018calibrated} for measuring how a recommendation algorithm is responsive to users' true preferences and we consider how various algorithms may result in different degrees of miscalibration. As we mentioned earlier, miscalibration, if it exists across all users, could simply mean failure of the algorithm to provide accurate  personalization. However, when different groups of users experience different levels of miscalibration, this could indicate unfair treatment of certain user groups.  From this standpoint, we call a recommender system unfair if it has different levels of miscalibration for different user groups.

In machine learning, a classification algorithm is called calibrated if the predicted proportions of the various classes agree with the actual proportions of classes in the training data. Extending this notion to recommender systems, a calibrated recommender system is one that reflects the various interests of a user in the recommended list, and with
their appropriate proportions. 

For measuring the miscalibration of the recommendations we use the metric introduced in \cite{steck2018calibrated}. Assume $u$ be a user and $i$ be an item. Also, suppose for each item $i$ there is a set of features $C$ describing the item. For example, a song could be Pop or Jazz, or a movie could have genres Action, Romance, Comedy, etc. We use $c$ for each of these individual categories. Also, we assume that each user has rated one or more items, showing interest in features $c$ belonging to those items. We consider two distributions of categories $c \in C$ for each user: one for the items rated by $u$, $P_u$, and the other one for all recommended items to $u$, $Q_u$.


For each feature $c$, we measure the ratio of items that have feature $c$ and rated by the user $u$, $p(c|u)$, and the ratio of items that have feature $c$ and are recommended to the user $u$, $q(c|u)$, as follows:

\begin{equation}
    p(c|u)=\frac{\sum_{i \in \Gamma_u}\mathbbm{1}(i \in c)}{ |\Gamma_u|} \;, \; \; \; \; \;  q(c|u)=\frac{\sum_{i \in \Lambda_u}\mathbbm{1}(i \in c)}{ |\Lambda_u|}
\end{equation}

where $\mathbbm{1}(.)$ is the indicator function returning zero when its argument is False and 1 otherwise. $\Gamma_u$ is the set of items rated by user $u$ and $\Lambda_u$ is the set of recommended items to user $u$. 

In order to determine if a recommendation list is calibrated to a given user, we need to measure the distance between the two probability distributions $P_u$ and $Q_u$.
We use the Hellinger distance, $H$, for measuring the statistical distance between these two distributions. We measure miscalibration for user $u$, $MC(P_u,Q_u)$, as follows:

\begin{equation}
    MC(P_u,Q_u)=H(P_u,Q_u)= \frac{\norm{\sqrt{P_u}-\sqrt{Q_u}}_2}{\sqrt{2}}  
\end{equation}

The overall miscalibration for each group $G$ is obtained by averaging $MC(P_u,Q_u)$
across all users $u$ in group $G$. 

\subsection{Popularity Bias}

Generally, rating data is skewed towards more popular items--there are a few popular items with the majority of the ratings while the rest of the items have far fewer ratings. Although it is true that popular items are popular for a reason, not every user has the same degree of interest towards these items \cite{oh2011novel,abdollahpouri2020multi}. There might be users who are interested in less popular, niche items. The recommender algorithm should be able to address the needs of those users as well.


Due to this common imbalance in the original rating data, often algorithms propagate and, in many cases, amplify the bias by over-recommending the popular items. In the next sections, we will define a metric for measuring the degree to which popularity bias is propagated / amplified by the recommendation algorithm. We will empirically evaluate how different recommendation algorithms propagate the popularity bias for different groups of users and to what extent it causes their recommendations to be miscalibrated.

\subsubsection{Algorithmic Popularity Lift}
In order to measure how each algorithm amplifies the popularity bias in its generated recommendations for different user groups, we define \textit{popularity lift}. First, we measure the average item popularity of a group \textit{G} (i.e. Group Average Popularity) as follows:

\begin{equation}
    GAP_p(G)=\frac{\sum_{u \in G} \frac{\sum_{i \in \Gamma_u}\theta(i)}{|\Gamma_u|}  }{|G|} \;, \; \; \; \; \;      GAP_q(G)=\frac{\sum_{u \in G} \frac{\sum_{i \in \Lambda}\theta(i)}{|\Lambda_u|}  }{|G|}
\end{equation}

where $\theta(i)$ is the popularity value for item $i$ (i.e. the ratio of users who rated that item) and subscript $p$ and $q$ refer to the profile of user and recommendations given to the user, respectively.

Therefore popularity lift ($PL$) for group $G$ is defined as:
\begin{equation}
    PL(G)=\frac{GAP_q(G)-GAP_p(G)}{GAP_p(G)}
\end{equation}

Positive values for $PL$ indicate amplification of popularity bias by the algorithm. A negative value for $PL$ happens when, on average, the recommendations are less concentrated on popular items than the users' profile. Moreover, the $PL$ value of 0 means there is no popularity bias amplification.

\section{Methodology}

We conducted our experiments on two publicly available datasets. The first one is MovieLens 1M dataset which contains 1,000,209 anonymous ratings of approximately 3,900 movies made by 6,040 users~\cite{movielens}. Each movie is associated with at least one genre in this dataset with a total of 18 unique genres in the entire dataset. The second dataset we used is a core-10 Yahoo Movies\footnote{https://webscope.sandbox.yahoo.com/catalog.php?datatype=r} which contains 173,676 ratings on 2,131 movies provided by 7,012 users. Similarly, in this dataset each movie is associated with at least one genre with a total of 24 genres in the entire datatset.

 For all experiments, we set aside a random selection of 80\% of the rating data as training set and the remaining 20\% as the test set. We used several recommendation algorithms including user-based collaborative filtering ($UserKNN$) \cite{resnick1997recommender}, item-based collaborative filtering ($ItemKNN$) \cite{sarwar2001item}, singular value decomposition (\textit{SVD++}) \cite{koren2008factorization}, and biased matrix factorization ($BMF$) \cite{koren2009matrix} to cover both neighborhood based and latent factor models. We also included the \textit{Most-popular} method (a non-personalized algorithm recommending the most popular items to every user) as an algorithm with extreme popularity bias. We tuned each algorithm to achieve its best performance in terms of precision. The precision values for \textit{UserKNN}, \textit{ItemKNN}, \textit{SVD++}, \textit{BMF}, and \textit{Most-popular} on MovieLens, are 0.214, 0.223, 0.122, 0.107, and 0.182, respectively. On Yahoo Movies, these values are 0.13, 0.127, 0.2, 0.047, and 0.1, respectively. We set the size of the generated recommendation list for each user to 10. We used the open source recommendation library \textit{librec-auto} \cite{mansoury2018automating} and LibRec 2.0 \cite{guo2015librec} for running our experiments.


In this paper, we are interested in seeing how different groups of users with varying degree of interest towards popular items are treated by the recommender system. Therefore, we grouped users in both datasets into an arbitrary number of groups (10 in this paper) based on their degree of interest in popular items. That is, we first measure the average popularity of the rated items in each user's profile and then organize them into 10 groups with the first group having the lowest average item popularity (extremely niche-focused users) and the last group with the highest average item popularity (heavily blockbuster-focused users). We denoted these groups as $G_1$ through $G_{10}$.

\section{Results}

\subsection{The Connection Between Popularity Bias and Miscalibration}\label{pop_bias_miscal}

In this section we show the connection between popularity bias and unfairness (disparate miscalibration) in recommendation. Making this connection would be helpful in many fairness-aware recommendation scenarios because fixing the popularity bias could be used as an approach to tackle this type of unfairness.

\begin{figure}
\centering
\SetFigLayout{2}{1}
  \subfigure[MovieLens]{\includegraphics[width=2.5in]{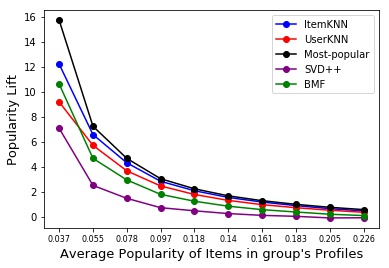}}
  \hfill
  \subfigure[Yahoo Movies]{\includegraphics[width=2.5in]{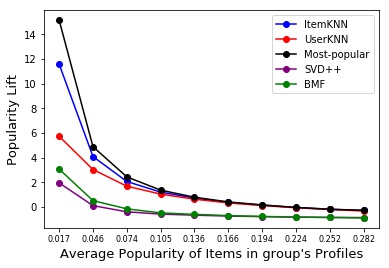}}
  \hfill
\caption{Average item popularity of user groups (G1 through G10 from left to right) and their observed popularity lift in two datasets.}
\label{pop_lift_pop}
\end{figure}

An illustration of the effect of the algorithmic popularity bias on different user groups is shown in Figure ~\ref{pop_lift_pop}. Each dot represents a group with certain average popularity of the users' profiles in that group which is shown on the x-axis. On the y-axis, the popularity lift of different algorithms on each user group is depicted. It can be seen that groups with the lowest average popularity (niche tastes) are being affected the most by the algorithmic popularity bias and the higher the average popularity of the group, the lesser the group is affected by the popularity bias. This shows how, unfairly, popularity bias is affecting different user groups.

Next, we want to show whether popularity lift has a connection with the miscalibration of the recommendations. Table ~\ref{table_pop_lift_extreme} shows the popularity lift and miscalibration experienced by two extreme groups: $G_1$ (the most niche-focused group) and $G_{10}$ (the most blockbuster-focused group). It can be seen that, for all of the algorithms, the popularity lift experienced by $G_1$ is significantly higher than $G_{10}$.
Also, it can be seen that, for all algorithms, the miscalibration value for $G_1$ is significantly higher than the one for $G_{10}$. This shows the group with the lowest average item popularity has experienced the highest popularity lift for their recommendations. Moreover, we also saw that this group experienced the highest miscalibration as well. This shows again how popularity lift might be associated with miscalibration.

\begin{table*}[t]
    \centering
    \small
    \caption{The popularity lift ($PL$) and Miscalibration ($MC$) of different recommendation algorithms on two groups $G_1$ and $G_{10}$. Bold values show significance difference with $p<0.05$.}
    \begin{tabular}{ l c c c c c c c c c c c}
        \toprule
        \multirow{3}{*}{algorithms} & \multicolumn{5}{c}{MovieLens} & & \multicolumn{5}{c}{Yahoo Movies} \\ \cline{2-6}\cline{8-12}
        & \multicolumn{2}{c}{$G_{10}$} && \multicolumn{2}{c}{$G_{1}$} && \multicolumn{2}{c}{$G_{10}$} && \multicolumn{2}{c}{$G_{1}$}  \\ \cline{2-3}\cline{5-6} \cline{8-9} \cline{11-12}
        & \thead{$PL$} & \thead{$MC$} && \thead{$PL$} & \thead{$MC$} && \thead{$PL$} & \thead{$MC$} && \thead{$PL$} & \thead{$MC$}  \\
            \bottomrule
        ItemKNN & 0.458 & (0.250) && \textbf{12.19} & \textbf{0.418} && -0.26 & 0.345 && \textbf{11.57} & \textbf{0.466}  \\
        UserKNN & 0.348 & 0.248 && \textbf{9.17} & \textbf{0.446} && -0.31 & 0.345 && \textbf{5.738} & \textbf{0.395} \\
        Most-popular & 0.563 & 0.277 && \textbf{15.7} & \textbf{0.501} && -0.25 & 0.342 && \textbf{15.13} & \textbf{0.471} \\
        SVD++ & -0.09 & 0.380 && \textbf{7.063} & \textbf{0.556} && -0.84 & 0.272 && \textbf{1.96} & \textbf{0.315} \\
        BMF & 0.086 & 0.396 && \textbf{10.60} & \textbf{0.635} && -0.871 & 0.290 && \textbf{3.079} & \textbf{0.345} \\
    \bottomrule
    \end{tabular}
\label{table_pop_lift_extreme}
\end{table*}

\subsection{Popular Item or Popular Feature?}

In Table \ref{table_pop_lift_extreme} we observed a connection between the degree of popularity lift experienced by a group of users and how miscalibrated their recommendations are. This connection can be justified as follows: when a list is miscalibrated, it means some genres (or features) are either over-represented or under-represented in the recommendation list compared to the users' interactions. Therefore, theoretically, it is possible for a list to be miscalibrated even when non-popular genres are over-represented in the recommendations. However, what happens in practice is that algorithmic popularity lift increases the recommendation frequency of popular movies and the genres associated with them. As a result, these popular genres become over-represented and thus causing overall miscalibration.

\begin{figure*}
    \centering
    \includegraphics[width=4.9in]{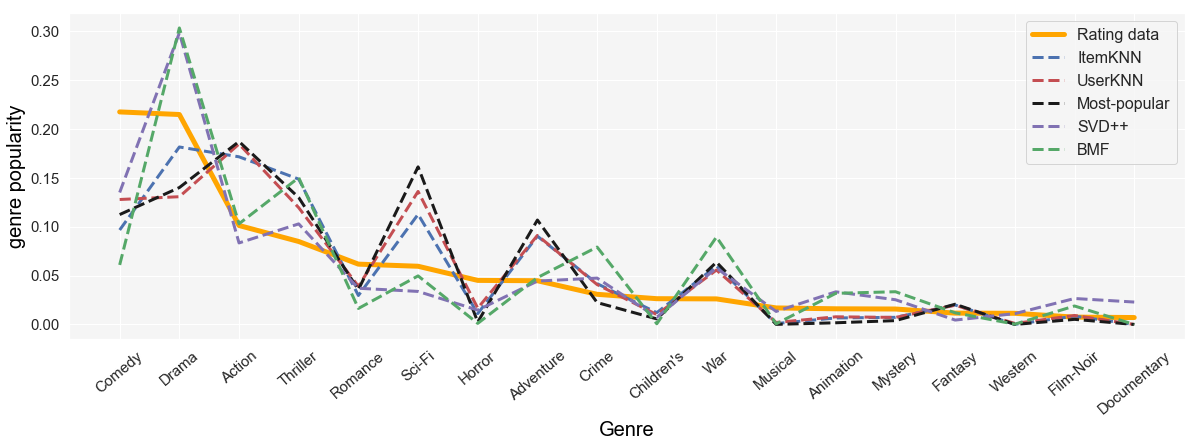}
    \caption{Genre popularity in rating data and in the recommendations \textbf{(MovieLens)}}
    \label{Genre_popularity_movielens}
\end{figure*}

\begin{figure*}
    \centering
    \includegraphics[width=4.9in]{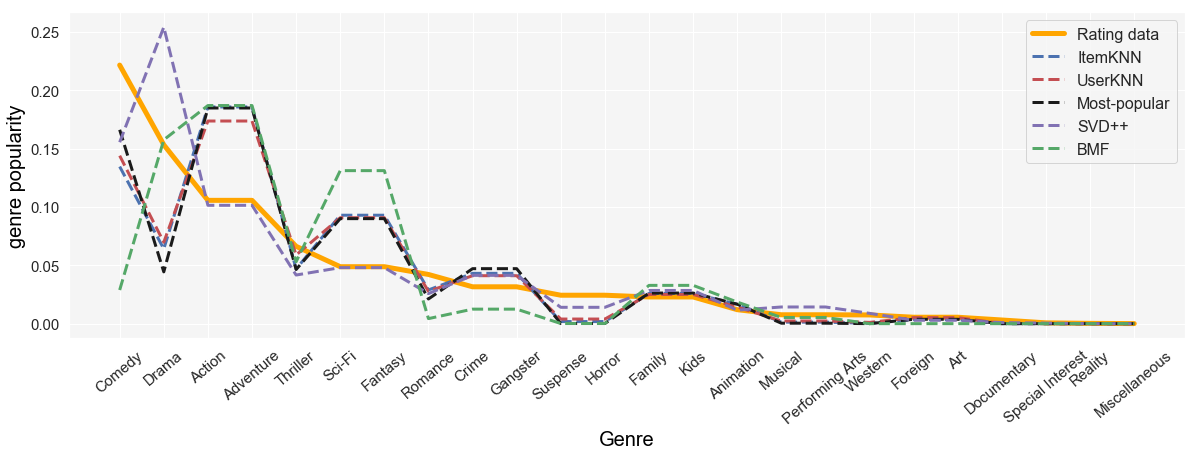}
    \caption{Genre popularity in rating data and in the recommendations \textbf{(Yahoo Movies)}}
    \label{Genre_popularity_yahoo}
\end{figure*}

Figure \ref{Genre_popularity_movielens} and \ref{Genre_popularity_yahoo} show the popularity (i.e. frequency) of different genres in rating data and in different recommendation algorithms in MovieLens and Yahoo Movies datasets, respectively. On both datasets, Comedy is the most popular genre. However, interestingly, the recommendations have not amplified this genre as one might expect. The reason is, the popularity of a genre does not necessarily mean every movies associated with that genre are also popular as it could simply be due to the fact that there are many movies with that particular genre. In fact, looking at the plot for MovieLens, we can see that genres Action, Thriller, Sci-Fi, and Adventure are amplified even though they are not the most popular genres. The reason is the most popular movies in this dataset are actually the ones that fall within those genres and since the recommendation algorithms are biased towards popular items (not genres) these genres are amplified. For instance, "Star Wars (episodes IV, V, and VI)", "Jurassic Park (1993)", "Terminator 2 : Judgment Day (1991)", and "The Matrix (1999)" are among the most popular movies and their associated genres are Action, Thriller, Adventure, and Sci-Fi. Similarly, on Yahoo Movies, Action, Adventure, Sci-Fi, Fantasy, Crime, and Gangster are amplified even though some of these genres are not the most popular ones in the rating data. The reason is, most popular movies in this data are "Pirates of the Caribbeans: The curse of the black pearl (2003)", "Terminators 3: Rise of the Machines (2003)", and "The Matrix reloaded (2003)" which are associated with Action, Adventure, Crime, Gangster, and Sci-Fi. This further supports our hypothesis that the popularity bias in recommendation could be a leading factor to miscalibration.

\section{Conclusion and Future Work}

In this paper, we looked at the popularity bias problem from the user's perspective and we observed different groups of users can be affected differently by this bias depending on how much they are interested in popular items. In particular, for two extreme groups on the spectrum of item popularity, we showed that the group which is less interested in popular items is affected the most by popularity bias and also has the highest level of miscalibration. 
For future work, we intend to further study the causality of popularity bias on miscalibration. In particular, we will design experiments such as sampling methods to control popularity bias in data and see the effect of that on miscalibration and fairness. We will also investigate how mitigating algorithmic popularity bias can help to lower miscalibration and unfairness.

\begin{acks}
Authors Abdollahpouri and Burke were supported in part by the \grantsponsor{NSF}{National Science Foundation}{} under Grant Number:~\grantnum{NSF}{1911025}.
\end{acks}



\bibliographystyle{ACM-Reference-Format}
\bibliography{main.bib}
\end{document}